# Examining the Legal Status of Digital Assets as Property: A Comparative Analysis of Jurisdictional Approaches

Luke Lee

King's College London

luke.lee@kcl.ac.uk

**Abstract**

This paper examines the complex legal landscape surrounding digital assets, analysing how they are defined and regulated as property across various jurisdictions. As digital assets such as cryptocurrencies and non-fungible tokens (NFTs) increasingly integrate with global economies, their intangible nature presents unique challenges to traditional property law concepts, necessitating a re-evaluation of legal definitions and ownership frameworks. This research presents a comparative analysis, reviewing how different legal systems classify and manage digital assets within property law, highlighting the variations in regulatory approaches and their implications on ownership, transfer, and inheritance rights. By examining seminal cases and regulatory developments in major jurisdictions, including the United States, the European Union, and Singapore, this paper explores the emerging trends and potential legal evolutions that could influence the global handling of digital assets. The study aims to contribute to the scholarly discourse by proposing a harmonized approach to digital asset regulation, seeking to balance innovation with legal certainty and consumer protection.

The emergence of digital assets represents a novel class of property within the rapidly evolving digital landscape. Their intangible essence challenges conventional property definitions and legal ownership principles, prompting a reassessment of property laws worldwide. Understanding the legal status of digital assets is crucial for effective regulation. Without clear classification and application of relevant legal principles, regulation risks ineffectiveness, exposing digital assets to potential misuse by malicious actors in cyberspace.

What the law cannot classify, it cannot regulate. As Haentjens and Lehmann aptly state, "*[n]otwithstanding the claims of some radical believers in the autonomy of the blockchain, digital assets are and will always be subject to the law and legal rules. Moreover, digital assets need law.*"[1] Accurate classification of digital assets is paramount as they become increasingly intertwined with the economy, impacting individuals, businesses, and governments alike. Unlike traditional assets, digital assets lack physical form, thereby raising questions about ownership, transfer, inheritance, and taxation. Thus, it is pivotal to decide whether to classify them as property and determine applicable property rights and to establish the applicable property rights.

## Chapter 1: The Concept and Types of Digital Assets

The definition of digital assets is subject to varying interpretations within academia. Austerberry defines them as any intellectual property or digital material with associated usage or distribution rights.[2] Conner broadens this definition to encompass any file created, stored, managed, or utilized through computers, servers, websites, or online platforms, including memberships and accounts.[3] The UCL Centre for Blockchain Technologies offers a more blockchain-centric approach, stating that digital assets are data recorded on a blockchain, conferring specific rights such as ownership, access, representation, voting, or practical use.[4] Underkuffler opines, "*The idea that property is 'things' is, however, easily discredited by lawyers and philosophers for its awkwardness and incompleteness.*"[5]

---

[1] Matthias Haentjens and Matthias Lehmann, 'Chapter 16 The Law Governing Secured Transactions in Digital Assets' in *Blockchain and Private International Law* (Brill Nijhoff, 2023).

[2] David Austerberry, *Digital Asset Management* (Taylor & Francis, 2013) as seen in Helen Akpan, Augustus Enyeribe and Awe, Michael, 'Digital Asset and PII Protection Using Blockchain Technology' (2022). Available from: doi:10.13140/RG.2.2.33569.48486.

At their core, digital assets consist of digital entities comprising data, content, and rights stored in a unique and identifiable manner. Technologies like blockchain facilitate this distinct identification, enabling their trade, usage, and ownership. The dynamic nature of property rights associated with digital assets reflects societal values and preferences in the digital era.[6] Whether digital assets possess property rights reflects our collective values and preferences in the digital age.[7] Therefore, understanding the nature of these rights is essential, with state legislatures encouraged to align regulatory approaches with broader societal values[8] rather than solely relying on private agreements.[9]

The scope of digital assets is extensive, covering a wide range of digitally stored content and rights[10] including cryptocurrencies, non-fungible tokens (NFTs), digital tokens, social network accounts, algorithms, cloud-based accounts, and other forms of digital intellectual property.[11,12] However, for the purposes of this academic discussion, the focus will be narrowed to the class of assets known as cryptoassets.

Cryptoassets (tokens or cryptocurrencies), as defined by the HMRC's Cryptoassets Manual, are digital assets secured through cryptography, representing value or contractual rights.[13] Sannitt further defines cryptoassets as "*representation of data on a distributed ledger cryptographically secured and updated subject to a decentralised consensus mechanism.*"[14] These assets are transferable, stored, and traded electronically.[15] While cryptoassets rely on Distributed Ledger Technology (DLT), not all DLT applications involve

---

[3] John Conner, 'Digital Life after Death: The Issue of Planning for a Person's Digital Assets after Death' (2011) 3 EPCPLJ 30, 301-21.

[4] UCL Centre for Blockchain Technologies, *Enterprise Digital Assets Report* (UCL Centre for Blockchain Technologies, 2022).

[5] Laura Underkuffler, 'The Idea of Property: Its Meaning and Power' SSRN as referenced in Natalie B Lynner, 'Property Interests in Digital Assets: The Rise of Digital Feudalism' (2017). 38 CLR 1099.

[6] Lynner (*n* 5).

[7] *Ibid*.

[8] *Ibid*.

[9] *Ibid*.

[10] UK Law Commission, 'Digital Assets: Final Report' (*UK Law Commission*, 27 June 2023). <https://lawcom.gov.uk/project/digital-assets/>, accessed on 20 February 2024.

[11] UK Law Commission (*n* 10).

[12] Akpan, Augustus and Michael (*n* 2).

[13] HM Revenue & Customs, 'Cryptoassets Manual' (*GOV.UK*, 21 August 2023) <https://www.gov.uk/hmrc-internal-manuals/cryptoassets-manual/crypto10100>, accessed on 20 February 2024.

[14] Adam Sanitt, 'What sort of property is a cryptoasset?' (Norton Rose Fulbright, February 2021) < https://www.nortonrosefulbright.com/en/knowledge/publications/26ade77a/what-sort-of-property-is-a-cryptoasset#:~:text=A%20cryptoasset%20also%20has%20an,term%20for%20the%20appropriate%20category.> accessed on 31 January 2024.

[15] HM Revenue & Customs (*n* 13).

cryptoassets.[16] Most commonly, digital assets, like cryptocurrencies, digital tokens, and NFTs, primarily use blockchain technology.

Blockchain technology is a subset of DLT,[17] facilitates decentralized, transparent, and intermediary-free transaction recording via cryptographic algorithms.[18,19] It enables the transfer of digital or digitally-represented physical assets without the need for trusted third parties,[20] ensuring accurate ownership records and fostering transparent and trustworthy information exchange.[21] Understanding the relationship between blockchain, cryptoassets, and smart contracts requires grasping its origins linked to Bitcoin, which serves as the cornerstone of its technological infrastructure.[22]

Bitcoin, authored by the enigmatic Satoshi Nakamoto, stands as the most widely recognized form of cryptoasset powered by blockchain technology. Nakamoto's seminal whitepaper titled "*Bitcoin: A Peer-to-Peer Electronic Cash System*"[23] laid the groundwork for Bitcoin, establishing it as the pioneering digital currency.[24] The whitepaper introduced groundbreaking innovations, including the decentralized consensus mechanism known as Proof-of-Work (PoW), which is crucial for securing and sustaining the network's operations.[25] Despite Bitcoin's prominence, the cryptocurrency landscape boasts over 4,000 to more than 7,800 diverse cryptocurrencies.[26] For instance, Ether (ETH) and Ripple (XRP) utilize different technologies from Bitcoin.[27] While sharing similarities with the Bitcoin blockchain, the Ethereum network's blocks contain transaction lists and the latest state, enhancing its decentralization and energy-efficient.[28] In 2022, Ethereum transitioned to a Proof of Stake consensus mechanism from PoW, further enhancing its efficiency.[29,30]

While, the Bitcoin blockchain laid the foundation for tokens, the true innovation arose with the integration of smart contracts and DLT.[31] Smart contracts automate

---

[16] *Ibid*.

[17] José M Garrido, *Digital Tokens: A Legal Perspective* (International Monetary Fund, 2023) 9.

[18] *Ibid*.

[19] Alexander Savelyev, 'Contract Law 2.0: 'Smart' Contracts as the Beginning of the End of Classic Contract Law' (2017) 26 ICTL 2, 119.

[20] *Ibid*.

[21] *Ibid*.

[22] *Ibid* 117.

[23] Satoshi Nakamoto, 'Bitcoin: A Peer-to-Peer Electronic Cash System' (2008) SSRN. Available from doi:10.2139/ssrn.3440802.

[24] UCL Centre for Blockchain Technologies (*n* 4) 24.

[25] *Ibid*.

[26] Gilead Cooper, 'Virtual Property: Trusts of Cryptocurrencies and Other Digital Assets' (2021) 27 T&T 625.

[27] *Ibid*.

[28] Vitalik Buterin, *A Next-Generation Smart Contract and Decentralized Application Platform* (Cm 37, 2014).

[29] Gaurav Roy, 'What is Ethereum Proof-of-Stake?' (*Ledger Academy*, 2 June 2023). <https://www.ledger.com/academy/ethereum-proof-of-stake-pos-explained>, accessed on 20 February 2024.

[30] Nico, 'Proof-of-Stake vs Proof-of-Work' (*Ethereum*, 25 January 2024) <https://ethereum.org/en/developers/docs/consensus-mechanisms/pos/pos-vs-pow/>, accessed on 20 February 2024.

[31] Garrido (*n* 17) 13.

interactions between parties[32] based on predefined rules encoded within the ledger. This facilitates the secure transfer and management of cryptoassets through blockchain-based smart contracts.[33] This development is significant as it enables the secure transfer and management of 'ownership' of cryptoassets through smart contracts on the blockchain.[34] However, this transferability of 'ownership' of the asset necessitates an examination of whether traditional property law principles remain applicable in a decentralised context.[35]

**Chapter 2: Legal Theories of Property and Their Application to Digital Assets**

Property has long been a cornerstone of legal frameworks, shaping economic and social structures across centuries. Traditionally, property encompasses both tangible and intangible assets, offering guidelines for ownership, transfers, and dispute resolution. Section 205 of the Law of Property Act 1925 defines property as "*anything in action, and any interest in real or personal property.*" However, digital assets present unique challenges to these traditional theories, prompting a re-evaluation of their applicability.

Traditional legal theories of property mainly focus on tangible assets, categorized into physicalist and bundle of rights theories. In Anglo-American law, property is classified as either a '*chose in possession*' or a '*chose in action*'[36,37,38] presenting a dilemma as traditional property must either be a tangible object or a legally enforceable right[39,40], neither of which fits cryptoassets.[41] The traditional approach under common law is as per *Colonial Bank v Whinney* (1885), wherein Fry LJ's famously opined "the law knows no *tertium quid* between the two [choses]".[42] In civil law systems following the Germanic tradition, Christian von Bar notes there is a tendency to narrowly define what constitutes *res incorporales*.[43] The question arises in civil law: Are crypto-asset rights treated as proprietary or contractual?[44]

Cryptoassets like Bitcoin challenge traditional theories of property. Bitcoin derives its value from perceived scarcity rather than physical attributes or contractual agreements,[45] defying conventional definitions of property. It neither constitutes a contractual right (*in personam*) nor exists as a physical object (*in rem*).[46] Traditional theories, like physicalist theory, assert that property must

---

[32] *Ibid*.

[33] *Ibid*.

[34] Rosa M Garcia-Teruel and Hector Simón-Moreno, 'The Digital Tokenization of Property Rights: A Comparative Perspective' (2021) 41 CLSR art 105543, 4.

[35] *Ibid*.

[36] *Colonial Bank v Whinney* (1885) 30 Ch D 261 (Court of Appeal) (Fry LJ).

[37] Cooper (*n* 26) 625.

[38] Garcia-Teruel and Simón-Moreno (*n* 34) 5.

[39] Cooper (*n* 26) 625.

[40] Garcia-Teruel and Simón-Moreno (*n* 34) 5.

[41] Cooper (*n* 26) 625.

[42] See *n* 36.

[43] Christian von Bar, (CH Beck 2015) 140-141, as cited in Jason Grant Allen, 'Cryptoassets in Private Law' in Iris Chiu and Gudula Deipenbrock (eds), *Routledge Handbook of Financial Technology and Law* (1st edn, Routledge 2021) 314.

[44] Jannik Woxholth and others, 'Competing Claims to Crypto-Assets' (2024) 28 ULR 226-246.

[45] *Ibid* 231.

[46] *Ibid*.

be tangible, leading to clear-cut ownership rights. However, the intangible nature of digital assets challenges this notion, as they cannot be possessed in the traditional sense. Garrido stated, "*Tokens connected to rights have immense potential, but also raise challenges from the legal point of view. The law's intervention is indispensable in establishing an enforceable link between tokens and the off-line reality, and in ensuring the correspondence between the tokens and the off-line assets or services.*"[47]

In the US, another theory, the 'bundle of rights' or 'bundle of sticks' theory, takes a more expansive approach, encompassing intangible assets.[48,49] In his famous 1960s essay on ownership, A. M. Honoré identified what he called the "*incidents of ownership*," now widely known as the 'bundle of rights'.[50] He argued these rights are consistent across all mature legal systems, though they can be modified by common law, statutes, or private agreements.[51] Key elements include the right to possess, use, manage, and earn income from property. These rights ensure that owners can control, benefit from, and ultimately dispose of their property, reflecting a framework adaptable to different legal contexts.[52] It conceptualizes property as a bundle of rights held by the owner, including the right to use, exclude others from use, and transfer ownership.[53] This 'bundle of rights' theory applies to both tangible and intangible properties, including intellectual property.[54]

Despite their replicability, digital assets challenge traditional notions of property rights due to their unique characteristics. Operating on decentralized networks like blockchain complicates control and jurisdiction over them, diverging from traditional property law's reliance on jurisdictional control. Digital assets often intersect with intellectual property rights but lack clear creators or owners, complicating their classification. Transferring and inheriting digital assets pose challenges unaddressed by traditional property laws. Globally, regulatory uncertainty persists, with diverse approaches ranging from acceptance to bans, further complicating their legal status. Technological advancement outpaces legal theory and regulatory development, leaving legal frameworks lagging behind.

While traditional property theories provide a foundational understanding, they struggle to adapt to the unique traits of digital assets. As digital assets gain prominence, the evolution of legal theories and regulations becomes essential to ensure clarity and security in this rapidly evolving realm.

**Chapter 3: Jurisdictional Analysis – Case Studies**

The classification of cryptoassets as property varies significantly across jurisdictions, influenced by differing legal traditions, regulatory priorities, and technological developments. A closer examination of Japan, UK, US, European

---

[47] Garrido (*n* 17) 5.

[48] Lior J Strahilevitz, 'Information Asymmetries and the Rights to Exclude' (2006) 104 MLR 1836.

[49] Denise R. Johnson, 'Reflections on the Bundle of Rights' (2007) 32 Vermont Law Review 247.

[50] A M Honoré, 'Ownership' in AG Guest (ed), Oxford Essays in Jurisprudence (1st edn, Oxford University Press 1961) 107-147.

[51] Johnson (*n* 49) 253.

[52] *Ibid.*

[53] *Ibid.*

[54] *Ibid* 247.

Union, Singapore, and China reveals diverse approaches and their implications.

Let us first start in Japan. On February 28, 2014, Mt. Gox, a Tokyo-based Bitcoin exchange, sought bankruptcy protection in Tokyo under a legal mechanism known as *minji saisei* (Kanji: 民事再生）, or civil rehabilitation pursuant to Japan's Civil Rehabilitation Act.[55][56] This process aimed to facilitate the court's search for a buyer by declaring liabilities of approximately 6.5 billion yen (US$65 million at that time) against assets worth 3.84 billion yen.[57] The MtGox insolvency case marked the first consideration of whether cryptoassets should be recognized as property. The Tokyo District Court's ruling in this case is notable for not conferring a proprietary nature to cryptoassets.[58]

The landmark judgment in *B2C2 Ltd v Quoine Pte Ltd* [2019] SGHC(I) 3 ("***B2C2***") by the Singapore International Commercial Court incorporated contract and trust law principles in a cryptocurrency trading case. Justice Simon Thorley (*obiter*) affirmed that "*Cryptocurrencies are not legal tender in the sense of being a regulated currency issued by a government but do have the fundamental characteristic of intangible property as being an identifiable thing of value.*"[59] The court concluded that cryptocurrencies meet traditional property rights criteria outlined by Lord Wilberforce,[60] including definability, identifiability, the capability of assumption by third parties, and a degree of permanence or stability.[61]

In *Vorotyntseva v Money-4 Limited t/a Nebeus.com & Ors* [2018] EWHC 2596 (Ch) ("***Vorotyntseva***"), issued in November 2019 following *B2C2*, claimant *Vorotyntseva* obtained a worldwide freezing order after depositing Ethereum and Bitcoin valued at £1.5 million with the Nebeus cryptocurrency platform.[62] Fearing the potential dissipation of her funds and lacking assurance from Nebeus regarding asset safety, she urgently sought the Court's intervention.[63] Justice Birss granted the freezing order due to the real risk of cryptoasset dissipation,[64] illustrating English courts' readiness to recognize cryptocurrencies as property.

---

[55] Ben McLannahan, 'Bitcoin exchange Mt Gox files for bankruptcy protection' (Financial Times, 1 March 2014) <https://www.ft.com/content/6636e0e8-a06e-11e3-a72c-00144feab7de> accessed 1 April 2024.

[56] Civil Rehabilitation Act (Act No. 225 of December 22, 1999). For an unofficial English translation of Japan's Civil Rehabilitation Act, refer to Civil Rehabilitation Act (Act No. 225 of 22 December 1999) (Japan Cabinet Secretariat) <https://www.cas.go.jp/jp/seisaku/hourei/data/cra.pdf> accessed 1 April 2024.

[57] Yoshifumi Takemoto and Sophie Knight, 'Mt. Gox files for bankruptcy, hit with lawsuit' (Reuters, 1 March 2014) <https://www.reuters.com/article/us-bitcoin-mtgox-bankruptcy-idUSBREA1R0FX20140228/> accessed 1 April 2024.

[58] Tokyo District Court, Reference number 25541521, Case claiming the bitcoin transfer etc., Tokyo District Court, Heisei 26 (Year of 2014), (Wa)33320, Judgement of Civil Division 28 of 5th August 2015 (Year of Heisei 27), Date of conclusion of oral argument; 10th June 2015, translation available at <https://www.law.ox.ac.uk/sites/files/oxlaw/mtgox_judgment_final.pdf.> accessed on 1 April 2024.

[59] *B2C2 Ltd v Quoine Pte Ltd* [2019] SGHC(l) 3 at [142].

[60] *National Provincial Bank v Ainsworth* [1965] 1 AC 1175 at 1247-8.

[61] *Ibid*.

[62] *Vorotyntseva v Money-4 Limited t/a Nebeus.com & Ors* [2018] EWHC 2596 (Ch).

[63] *Ibid*.

[64] *Ibid*.

Subsequent to *Vorotyntseva*, on 18 November 2019, the UK Jurisdiction Taskforce (UKJT)[65] published the Legal Statement on Cryptoassets and smart contracts regarding the legal status of cryptoassets and smart contracts under English law ("**Legal Statement**"). Operating under LawtechUK, UKJT comprises industry leaders aiming to support the digital transformation of the UK's legal services sector and position English law as a preferred option for emerging technologies. The Legal Statement, described by UKJT Chair and Chancellor of the High Court, Sir Geoffrey Vos as "*a watershed for English law and the UK's jurisdictions. Our statement on the legal status of cryptoassets and smart contracts is something that no other jurisdiction has attempted*," profoundly impacted jurisprudence following its publication.

The implications of these cases and legal statements underscore the evolving landscape of cryptoasset regulation and the importance of legal clarity in navigating this complex terrain.

**After the Legal Statement**

*AA v Persons Unknown* [2019] EWHC 3556 (Comm) ("*AA*") stands as a landmark case in the UK High Court, examining whether Bitcoin constitutes property. Although not the first of its kind, *AA* is notable for being the first detailed examination and reported case following the Legal Statement. [66] In *AA*, the Court heavily relied on the Legal Statement, with the Justice Bryan stating as follows: "*in my judgment, it is relevant to consider the analysis in that Legal Statement as to the proprietary status of cryptocurrencies because it is a detailed and careful consideration and, as I shall come on to, I consider that that analysis as to the proprietary status of cryptocurrencies is compelling and for the reasons identified therein should be adopted by this court*."[67] In *AA*, the Justice Bryan noted that the Legal Statement is misnamed, clarifying that it lacks binding legal effect as it "*is not in fact a statement of the law*."[68] He heavily relied on the reasoning from the Legal Statement to conclude that "*for the reasons identified in that legal statement, I consider that crypto assets such as Bitcoin are property*."

In the United States, digital assets are primarily regulated as securities or commodities based on their characteristics and usage. The Securities and Exchange Commission (SEC) typically treats digital assets resembling traditional investments as securities, while the Commodity Futures Trading Commission (CFTC) classifies cryptocurrencies like Bitcoin as commodities. Key legal cases, such as *SEC v. W.J. Howey Co.*, have played a crucial role in determining the application of securities laws to digital assets, employing the "Howey Test" to evaluate investment contracts. Both the US and EU prioritize incorporating digital assets into existing legal and regulatory frameworks, focusing on market stability and investor protection, albeit potentially hindering innovation. As major economic powers, their approaches significantly influence global digital asset markets, emphasizing consumer protection in the EU and US, which may dampen market risks but impede rapid development. Ongoing legal cases and SEC decisions in the US contribute to occasional uncertainty in the regulatory landscape.

---

[65] UK Jurisdiction Taskforce, 'Legal Statement on Cryptoassets and Smart Contracts' (*UK Jurisdiction Taskforce,* 2019) <https://technation.io/wpcontent/uploads/2019/11/6.6056_JO_Cryptocurrencies_Statement_FINAL_WEB_111119-1.pdf>, accessed on 20 February 2024.

[66] Vicki Ball, 'Bitcoin as Property: AA v Persons Unknown Re Bitcoin' (2020) 2 CPPL 186-94.

[67] *AA v Persons Unknown* [2019] EWHC 3556 (Comm) at [57].

[68] *Ibid*.

The European Union aims to standardize the classification and regulation of digital assets through the Markets in Crypto Assets Regulation (MiCA).[69] The European Parliament passed the final version of the text in a plenary session on April 20, and the official signing took place on May 31, 2023.[70] This regulation is set to be enforced starting December 30, 2024.[71] MiCA amends Regulations (EU) No 1093/2010 and (EU) No 1095/2010 and Directives 2013/36/EU and (EU) 2019/1937 and introduces unified rules for crypto-assets across the EU.[72] This legislation ensures legal clarity for crypto-assets that are not covered by existing EU laws, enhancing consumer and investor protection and financial stability.[73]

MiCA encourages innovation and the adoption of crypto-assets by defining three specific categories: asset-referenced tokens (ART), electronic money tokens (EMT), and other crypto-assets.[74] Additionally, the regulation governs the issuance, trading, and asset management of these crypto-assets, with particular focus on significant ART and EMT. Under Article 143 of MiCA, member states are permitted to implement 'transitional measures' that allow entities already engaged in providing crypto-asset services under national laws to continue operating during the 18-month transitional period following MiCA's full implementation in December 2024.[75] These measures include a 'grandfathering' clause under Article 143(3), which permits entities providing crypto-asset services before December 30, 2024, to continue until July 1, 2026, or until they are either granted or denied MiCA authorization.[76] Additionally, a simplified authorization process (Article 143(6)) is available for entities that were already authorized under national laws as of December 30, 2024, to provide crypto-asset services.[77]

In *Armstrong v Winnington* [2012] EWHC 10, an EU carbon emissions allowance was considered intangible property. The EU's MiCA, once implemented, could provide more clarity and consistency across member states. Executive Vice President of the European Commission for An Economy that Works for People stated, "*Lack of legal certainty is often cited as the main barrier to developing a sound crypto-asset market in the EU. This is a good chance for Europe to strengthen its international standing and become a global standard-setter with*

---

[69] European Parliamentary Research Service, Issam Hallak, 'EU Legislation in Progress: Markets in crypto-assets (MiCA)' (European Parliament, 2022) <https://www.europarl.europa.eu/RegData/etudes/BRIE/2022/739221/EPRS_BRI(2022)739221_EN.pdf> accessed 1 April 2024.

[70] European Parliamentary Research Service (*n* 69) 9.

[71] *Ibid.*

[72] European Parliament, 'Markets in crypto-assets (MiCA)' (European Parliamentary Research Service, 29 September 2023) https://www.europarl.europa.eu/thinktank/en/document/EPRS_BRI(2022)739221 accessed 1 April 2024.

[73] European Parliament (*n* 72).

[74] *Ibid.*

[75] European Securities and Markets Authority, 'Markets in Crypto-Assets Regulation (MiCA)' <https://www.esma.europa.eu/esmas-activities/digital-finance-and-innovation/markets-crypto-assets-regulation-mica> accessed 1 April 2024.

[76] *Ibid.*

[77] *Ibid.*

*European companies leading new technologies for digital finance.*"[78]

Singapore's progressive stance on digital assets is highlighted by the Monetary Authority of Singapore (MAS)'s nuanced approach of regulating with flexibility. MAS classifies digital assets based on their characteristics, either as property or securities. Singapore's Payment Services Act 2019 ("**PSA**"), passed by the Singapore Parliament on 14 January 2019[79], provides a regulatory framework for digital payment token services. The PSA consolidates various earlier legislative efforts on payment services that had not been designed with the present FinTech innovations in mind. One laudable effort by the PSA is its clear definition of virtual currencies and cryptocurrencies as "*digital payment tokens*" (DPTs).[80] DPTs, as per section 2 of the PSA, are "*any digital representation of value (other than an excluded digital representation of value) that —*

*(a) is expressed as a unit;*

*(b) is not denominated in any currency, and is not pegged by its issuer to any currency;*

*(c) is, or is intended to be, a medium of exchange accepted by the public, or a section of the public, as payment for goods or services or for the discharge of a debt;*

*(d) can be transferred, stored or traded electronically; and*

*(e) satisfies such other characteristics as the Authority may prescribe;*"[81]

The Monetary Authority of Singapore (MAS) describes the PSA as an "*omnibus framework*", which encompasses both modern and traditional payment methods.[82] This balanced approach supports Singapore's regulatory oversight by safeguarding consumers of DPTs while fostering a pro-business environment. In *ByBit Fintech Ltd v Ho Kai Xin and others* [2023] SGHC 199 ("*ByBit*"), the Singapore High Court ruled that cryptoassets can be considered property held in trust, a significant development in Singapore's digital asset jurisprudence.[83] Singapore's flexibility, adaptability and progressive policies therefore position it as a hub for digital asset innovation.

China takes a more restrictive stance on digital assets, especially cryptocurrencies, with the People's Bank of China (PBoC) banning all cryptocurrency transactions and initial coin offerings (ICOs) in late September 2021.[84] Instead, China focuses on developing its state-controlled digital currency, the Digital Currency Electronic Payment (DCEP)[85], prioritizing financial stability and control.

---

[78] European Commission, 'Digital Finance: Commission Holds Closing Pan-European Conference Following Extensive Outreach Events' (*European Commission*, 23 June 2020) <https://ec.europa.eu/commission/presscorner/detail/en/ip_20_1136>, accessed on 20 February 2024.

[79] Singapore, Payment Services Act 2019 (Act 2 of 2019) <https://sso.agc.gov.sg/Acts-Supp/2-2019/Published/20190220?DocDate=20190220> accessed 1 April 2024.

[80] *Ibid.*

[81] *Ibid.*

[82] Monetary Authority of Singapore, 'A Guide to The New Payment Services Act 2019' <https://www.mas.gov.sg/-/media/MAS/FAQ/Payment-Services-Act-Infographic> accessed 1 April 2024.

[83] *ByBit Fintech Ltd v Ho Kai Xin and others* [2023] SGHC 199.

[84] Francis Shin, 'What's behind China's cryptocurrency ban?' (World Economic Forum, 31 January 2022) <https://www.weforum.org/agenda/2022/01/what-s-behind-china-s-cryptocurrency-ban/#:~:text=In%20late%20September%202021%2C%20the,to%20their%20highly%20speculative%20nature> accessed 1 April 2024.

While some jurisdictions recognize digital assets as property with clear legal frameworks, others struggle to categorize them, leading to regulatory uncertainty. Technological advancements outpace legal adaptation,[86] but significant legal developments are underway as courts and legislatures apply established principles to innovative scenarios.[87] The diverse approaches worldwide reflect ongoing legal evolution in response to digital asset challenges and opportunities.

## Chapter 4: Impact on Stakeholders and Markets

The legal status of digital assets significantly influences stakeholders, comprising investors, companies, and regulators, thereby shaping market dynamics and carrying legal and financial risks. The recognition of digital assets provides investors with clarity and security, encouraging market participation and ensuring consumer protection. Conversely, regulatory uncertainty deters investment, leading to market volatility and hindering mainstream adoption.

For companies, legal acknowledgment legitimizes their operations, driving innovation and fostering partnerships. However, lacking clear legal status, they encounter operational uncertainties and difficulties accessing financial services. Regulatory bodies depend on legal recognition to effectively monitor and control digital asset activities, preserving market integrity and deterring illicit actions.

Well-defined legal frameworks typically boost investor confidence, prompting increased investment in digital assets and catalysing innovation and sectoral growth. Conversely, legal ambiguity fosters cautious investment strategies, constraining market expansion. Market stability thrives on clear legal rules, attracting more participants and fueling growth, while uncertainty can induce volatility and impede long-term development.

Robust legal and regulatory frameworks expedite the integration of digital assets into mainstream financial systems, enhancing their utility and enlarging their user base. Nevertheless, uncertain legal status exposes companies and investors to compliance risks, potentially resulting in legal breaches and penalties. Market integrity faces threats without stringent frameworks, heightening susceptibility to manipulation and fraud.

Operational risks for digital asset companies include challenges in securing banking services and enforcing contracts. Taxation and financial reporting complexities further compound risks for investors and companies. Ambiguous legal status elevates the risk of litigation pertaining to asset ownership, transfer, and valuation disputes.

---

[85] Arjun Kharpal, 'China has given away millions in its digital yuan trials. This is how it works' (CNBC, 4 March 2021) <https://www.cnbc.com/2021/03/05/chinas-digital-yuan-what-is-it-and-how-does-it-work.html> accessed 1 April 2024.

[86] Michael Sinclair and Sophy Teng, 'Singapore Court's Cryptocurrency Decision: Implications for Cryptocurrency Trading, Smart Contracts and AI' (*Norton Rose Fulbright*, September 2019) <https://www.nortonrosefulbright.com/en-gb/knowledge/publications/6a118f69/singapore-courts-cryptocurrency-decision-implications-for-trading-smart-contracts-and-ai#17> accessed on 20 January 2024.

[87] *Ibid*.

The legal recognition of digital assets carries profound implications for stakeholders and market dynamics. Such recognition fosters clarity, security, and stability, encouraging investment and innovation. Conversely, the absence of recognition precipitates uncertainties, market volatility, and heightened legal and financial risks.

As the market progresses, clear legal frameworks will be instrumental in mitigating risks, promoting stability, and safeguarding stakeholders.

## Chapter 5: Future Trends and Possible Legal Developments

There is a growing recognition of the need for clarity and certainty in the legal status of digital assets as the landscape continues to evolve. Savelyev advocates for a distinct *sui generis* legal framework tailored to the unique characteristics of digital assets.[88] However, given the ongoing development and uncertainties surrounding these assets, comprehensive regulations may be premature.[89] Instead, integrating them into existing legal frameworks, both directly and by analogy, remains essential in the interim.[90]

Over time, specific legal rights for tokens may develop, drawing from securities law, tangible property rights, and intellectual property rights.[91] A robust legal framework is essential to recognize the benefits of DLT and support continuous technological advancements while addressing unresolved issues.[92] Woxholth et al. propose a strategy based on three principles: recognizing ownership rights in cryptoassets, connecting protection for genuine purchasers, and facilitating enforcement of rights concerning cryptoassets.[93] They advocate for legislative measures to assign property rights, safeguard *bona fide* purchasers, and streamline law enforcement efforts.[94] This involves licensing requirements for asset custodians, compliance with Know Your Customer (KYC) and tracking regulations, and legal actions against digital asset custodians.[95] These actions aim to ensure data availability for legal disputes and facilitate the resolution of competing claims, particularly in bankruptcy cases.[96]

International collaboration, tech-neutral legislation, flexible frameworks, and stakeholder involvement are recommended for harmonizing legal approaches. Unclear regulations can lead to costly legal battles and jurisdictional conflicts,[97] while some nations may aim to attract cryptocurrency businesses by becoming "free havens," potentially undermining law enforcement efforts.[98] Blockchain networks implementing terms of use favouring *laissez-faire* jurisdictions may result in legal disputes over public order restrictions.[99] Achieving complete uniformity in national private laws may be unrealistic, but an international consensus

---

[88] Savelyev (*n* 19) 869.

[89] *Ibid*.

[90] *Ibid*.

[91] *Ibid*.

[92] Woxholth and others (*n* 44) 240.
[93] *Ibid*.

[94] *Ibid* 246.

[95] *Ibid*.

[96] *Ibid*.

[97] *Ibid* 245.

[98] *Ibid*.

[99] *Ibid*.

on certain levels could enhance the effectiveness of regulatory efforts.[100]

The evolving regulatory landscape of digital asset regulation underscores their increasing importance in the global economy. There is a trend towards more sophisticated legal frameworks to tackle the unique challenges posed by digital assets. Many jurisdictions are moving towards clearer regulatory frameworks, which are crucial for ensuring certainty in token operations. This involves addressing issues like transfer rules, security interests, insolvency effects, and jurisdictional matters.[101,102] Clear definitions of tokens under securities laws are vital to avoid stifling innovation.[103,104] A strong legal foundation supports technological advancement, offering economic benefits like liquidity, cost savings, and security.[105] Legislation should prioritize creating structures that foster innovation and economic deployment.[106]

Defining the legal status of digital assets like cryptocurrencies, NFTs, and digital tokens requires establishing regulations for their use, taxation, and transfer, with a growing emphasis on consumer protection. Regulations aim to prevent fraud, ensure transparency, and combat market manipulation, supported by strict anti-money laundering (AML) and know-your-customer (KYC) policies. Digital assets are increasingly integrated into traditional financial systems, prompting considerations from central banks and inclusion in mainstream financial products and services.

Legal theories regarding property may evolve to encompass digital assets more comprehensively, potentially introducing new categories that acknowledge their unique nature. International law may develop to address cross-border issues, harmonizing regulations, and enforcement. Intellectual property law may also see significant changes in how rights are assigned, transferred, and enforced digitally.

Nations should collaborate to establish common standards and frameworks for digital asset regulation, possibly through international agreements or treaties. Laws should be technology-neutral, focusing on function and use rather than underlying technology to remain relevant amid rapid advancements. Flexible legal frameworks that adapt to new developments are crucial, with regulators engaging various stakeholders to ensure well-informed and balanced legislation.

Introducing security tokens might enable regulators to adopt a more proactive stance on compliance, highlighting an intriguing interplay between law and technology.[107] Common law systems in countries like Canada, the US, the UK, and Australia rely on legal precedent, evolving gradually through past case outcomes.[108] However, this retrospective nature and the lengthy

---

[100] *Ibid*.

[101] Garrido (*n* 17) 5.

[102] *Ibid*.

[103] *Ibid*.

[104] *Ibid*.

[105] *Ibid*.

[106] *Ibid*.

[107] Brian Leiberman and Dave Mirynech, *Digital Assets: The Era of Tokenized Securities* (Cutting-Edge Innovation in the Cryptosphere, 2019).

[108] *Ibid*.

legal process often lags in adapting to technological advancements.

Contrastingly, technological innovation is marked by relentless forward momentum, a concept Ray Kurzweil, Google's Chief Futurist and Director of Engineering, describes as the "*Law of Accelerating Returns.*"[109] This disparity is evident in the digital asset realm,[110] where the SEC applied outdated securities laws following the 2017 ICO boom.[111] Yet, the customizable nature of tokenized securities offers regulators an opportunity to embed regulations directly into token issuance, enabling more immediate and proactive enforcement.[112]

The future of digital asset regulation and legal theory will likely emphasize increased clarity, consumer protection, and international cooperation. Legal frameworks are expected to evolve to accommodate the unique characteristics of digital assets, with a focus on harmonizing approaches across jurisdictions to manage their global nature effectively. Gilead Cooper KC suggests, "[i]*t may be that legislation will ultimately be needed to give a coherent structure to the law of what is, in effect, an entirely new kind of property: A Law of Virtual Property Act?*"[113]

## Chapter 6: Conclusion

In conclusion, the examination of digital assets across legal and jurisdictional lenses underscores the intricate complexity and ever-evolving nature of their classification and regulation. It is evident that while jurisdictions like the United States and the European Union are striving to integrate digital assets within existing legal frameworks, others such as Singapore are adopting more flexible approaches, while China is opting for stringent controls. This diversity reflects varying regulatory philosophies, the delicate balance between innovation and stability, and the challenges of harmonizing investor protection with market development.

The determination of the legal status of digital assets holds immense significance, with far-reaching implications for stakeholders, market dynamics, and the broader financial landscape. Whether digital assets receive legal recognition or not shapes investor confidence, market stability, and the pace of innovation. Moreover, the emerging trends and potential developments in legal theory surrounding digital assets underscore the necessity for adaptive and forward-thinking legal frameworks.

Looking ahead, the legal landscape for digital assets is poised for further evolution. The crux lies in crafting regulations that can keep pace with rapid technological advancements while safeguarding market integrity and protecting stakeholders' interests. Given that digital assets continuously blur traditional legal boundaries, the need for international collaboration and harmonized legal approaches is becoming increasingly critical. The future legal framework for digital assets will likely entail a delicate balance between fostering innovation and ensuring regulatory compliance and stability. Evolving legal frameworks are essential to address the inherent complexities of digital assets, with international collaboration and flexible legal strategies taking on greater importance to provide a solid regulatory foundation for the future.

---

[109] *Ibid*.

[110] *Ibid*.

[111] *Ibid*.

[112] *Ibid*.

[113] Cooper (*n* 26) 631.

The classification of digital assets as property, as it determines the applicability of property laws, including rights of ownership, transferability, and enforceability of those rights in courts of law. Furthermore, the legal recognition of digital assets as property carries significant implications for investor protection, market stability, and the development of regulatory frameworks that can adapt to technological advancements. This classification impacts the extent of rights and protections for investors and asset holders, influences operational frameworks for businesses—especially in sectors like finance, technology, and media—and shapes the development and enforcement of policies and regulations by governments and regulatory bodies, ultimately impacting national and global economies.

**Acknowledgements:**

This working paper authored by Luke Lee expands upon research originally conducted as part of the academic requirements for the Master of Laws (LL.M) degree at The Dickson Poon School of Law, King's College London, during the academic year 2024. A version of this work was submitted in fulfilment of the LL.M degree. This paper aims to contribute to the scholarly discourse on the law of cryptocurrencies and digital asset regulation. Special thanks are extended to Professor Bernhard Maier and Visiting Lecturer Penelope Nevill for their invaluable academic feedback on the research plan for the original submission. Save for the aforementioned feedback, no external supervision, funding, or ethics review was required for the development of both the original submission and this expanded research.